\documentclass[journal=eds]{CUP-JNL-DTM}%

\usepackage[normalem]{ulem}
\usepackage{graphicx}
\usepackage{multicol,multirow}
\usepackage{amsmath,amssymb,amsfonts}
\usepackage{mathrsfs}
\usepackage{amsthm}
\usepackage{rotating}
\usepackage{appendix}
\usepackage{ifpdf}
\usepackage[T1]{fontenc}
\usepackage{newtxtext}
\usepackage{newtxmath}
\usepackage{textcomp}
\usepackage{xcolor}
\usepackage{lipsum}
\usepackage[colorlinks,allcolors=blue]{hyperref}

\usepackage{subcaption}
\usepackage{caption}
\usepackage[labelfont=bf]{caption}

\theoremstyle{definition}

\numberwithin{equation}{section}

\jname{Environmental Data Science}
\articletype{APPLICATION PAPER}
\jyear{2025}

\begin{document}

\begin{Frontmatter}

\title[Article Title]{Optimal sensor placement for the reconstruction of ocean states using differentiable Gumbel-Softmax sampling operator}

\author[1,2]{Oscar Chapron}
\author[1,2]{Ronan Fablet}
\author[3]{Yann Stéphan}

\address[1]{\orgname{IMT Atlantique}, \orgaddress{\city{Brest}, \country{France}}. \email{oscar.chapron@imt-atlantique.fr}}
\address[2]{\orgname{Lab-STICC}, \orgaddress{\city{Brest}, \country{France}}}
\address[3]{\orgname{SHOM}, \orgaddress{\city{Brest}, \country{France}}}
\authormark{Oscar Chapron \textit{et al}.}

\keywords{Adaptive Sensor Placement, Gumbel-Softmax, End-to-End Optimization, Ocean State Estimation}

\abstract{Accurately reconstructing and forecasting ocean fields from sparse observations is critical for both operational and scientific purposes. Optimizing sensor placement to maximize reconstruction skill remains challenging due to evolving ocean dynamics and practical deployment constraints. Traditional approaches, such as Empirical Orthogonal Functions, greedy search, or Gaussian processes, either assume static observation networks or scale poorly in high-resolution and non-stationary regimes.

We introduce a differentiable adaptive sensor placement framework based on a Gumbel-Softmax sampling operator. Given an ensemble of forecasts or simulations, the method jointly optimizes a probabilistic sampling mask and the reconstruction mapping (e.g., Optimal Interpolation correlation lengths) under strict observation budgets. Numerical experiments are conducted for Sea Surface Height reconstruction in a Gulf Stream region through Observing-System Simulation Experiments using a state-of-the-art high-resolution ocean simulation.

With a sensor budget of only 0.1\% (fewer than 100 point-wise observations on a 14°×14° domain) the optimized sampling reduces the reconstruction RMSE by more than half (0.0908 m versus 0.1750 m) and increases explained variance by about 20\% (93.1\% versus 74.4\%) compared with a uniform random strategy. The method remains robust when trained on noisy ensembles with controlled location uncertainty errors (up to 1°), supporting robustness under the forecast-uncertainty assumptions considered here.

Overall, the framework provides a scalable, budget-aware approach to designing observation networks. Beyond improved skill, it yields interpretable sampling patterns that consistently target energetic regions such as eddies and fronts, offering a transferable tool for adaptive sensing in geophysical systems.}

\end{Frontmatter}

\section*{Impact Statement}
This study introduces a scalable, differentiable framework for adaptive ocean sensor placement that jointly optimizes where to observe and how to reconstruct dynamic sea surface fields under strict budget constraints. By leveraging Gumbel-Softmax sampling and Optimal Interpolation, the method dramatically improves reconstruction accuracy while maintaining robustness to ensemble forecast errors. Beyond outperforming static and random baselines, the learned sampling patterns are interpretable and align with key oceanographic structures, offering a promising methodological step toward, data-driven observation networks in climate and Earth system science.

\section{Introduction}
Ocean dynamics, ranging from small-scale turbulence and mesoscale eddies to basin-wide currents and global circulation patterns like the Atlantic Meridional Overturning Circulation (AMOC), play a pivotal role in regulating Earth’s climate, marine ecosystems, and coastal resilience. Accurate monitoring of key ocean variables, such as sea surface height (SSH), is thus critical for improving weather and ocean forecasts, and mitigating risks associated with sea-level rise and extreme events. Yet, despite advances in space oceanography and in situ platforms like Argo floats, the observational coverage remains sparse and irregular, particularly in remote regions. This sparsity poses a fundamental challenge in reconstructing high-fidelity gap-free estimates of ocean variables from limited observations while operating under limited sensor deployment budgets.

Observation-based reconstruction methods, such as Optimal Interpolation (OI) (\cite{daley1993atmospheric,taburet2019duacs}), leverage spatial covariance structures to interpolate observations onto regular grids. While computationally efficient, OI assumes stationary correlation scales and Gaussian statistics, limiting its ability to resolve non-linear features like sharp frontal zones or eddies—phenomena critical to ocean energy transfers and cycles. Model-based reconstruction methods based on data assimilation schemes offer a well-posed inverse problem formulation to combine observation data and physical models. While they deliver the state-of-the-art performance for the forecasts and reanalyses of the full ocean states (\cite{fablet2021learning, martin2023synthesizing}), they still face challenges in actually exploiting the potential of observational datasets compared with observation-based reconstruction methods. This is particularly highlighted for the reconstruction of sea surface variables such as SSH (\cite{ballarotta2020dynamic}). 
Recently, deep learning approaches (\cite{fablet2021learning, martin2023synthesizing}) have shown promise in learning complex data-driven mappings from sparse samples to gap-free fields.
Their performance remains tightly coupled to the spatial distribution of observations. As shown in (\cite{krause2008near}), suboptimal sensor placement can degrade reconstruction accuracy by up to 40\% in data-sparse regimes, underscoring the need for principled strategies to optimize observational resource allocation.

Adaptive sensor placement aims to strategically position the sensors to maximize the information gain (\cite{manohar2018data, ma2025physense, wang2023learning}). The problem is fundamentally linked to Optimal Experimental Design (OED) theory (\cite{pukelsheim2006optimal,papalambros2000principles}), which provides formal criteria for assessing the quality of sensor configurations. The three most commonly used OED criteria are: A-optimality, which minimizes the average estimation variance; D-optimality, which maximizes the overall information content (or determinant of the information matrix); and E-optimality, which minimizes the worst-case estimation error. In this study, we primarily adopt the A-optimality criterion. Although these principles originated in classical OED, they have been extended to the sensor placement problem by integrating spatial and physical constraints (\cite{krause2008near}). It offers a pathway to address the above limitations. These approaches scale poorly with grid resolution and lack in general integration with state-of-the-art reconstruction schemes. Machine learning breakthroughs, particularly in differentiable sampling (\cite{balin2019concrete}), now enable end-to-end optimization of sensor placement while differentiating through the reconstruction process. The Gumbel-Softmax trick (\cite{jang2016categorical, maddison2016concrete}), for instance, provides a continuous relaxation of discrete sampling decisions, allowing gradient-based training of probabilistic sensor masks under budget constraints. This paradigm shift opens new opportunities to co-design sensing and reconstruction systems tailored to specific oceanographic features.

In this work, we explore this approach and address optimal sensor placement for the reconstruction of sea surface dynamics from sparse observations. We propose a novel framework for adaptive sensor placement. It combines the probabilistic flexibility of a Gumbel-Softmax sampling with the interpretability of Optimal Interpolation. Our approach jointly optimizes sensor locations and OI covariance parameters with a view to automatically prioritizing regions of high dynamical interest, such as eddy-rich zones or boundary currents. Our main contributions are three-fold

\textbf{Key Contributions}
\begin{enumerate}
    \item A Differentiable Probabilistic Sensor Placement Model: we introduce a Gumbel-Softmax-derived sampling operator with location-specific sampling probabilities. The Gumbel-Softmax trick (\cite{jang2016categorical, maddison2016concrete}) ensures the full differentiability of the sampling operator with respect to  model parameters.
   \item  An End-to-end Gradient-Based Co-Design : An end-to-end optimization framework that jointly optimizes the sensor sampling operator and reconstruction parameters (e.g., OI correlation lengths) by minimizing a physics-informed reconstruction loss under strict budget constraints. 
   \item An application to in situ observation networks of sea surface dynamics: in a simulation-based case study, we demonstrate how the proposed approach optimizes the placement of in situ sensors given ensemble forecasts for the reconstruction of SSH in a highly-dynamic Gulf Stream region, with a 48\% reduction in mean squared error over baseline methods.
\end{enumerate}

The remainder of this paper is organized as follows. Section 2 reviews the related work in sensor placement in environment monitoring. Section 3 details the proposed methodology, including the Gumbel-Softmax formulation and its integration in an OI framework. Section 4 presents our numerical experiments on SSH, and Section 5 discusses our main findings and future research directions.

\section{Problem statement and Related works}

The challenge of optimal sensor placement has been addressed through multiple perspectives across geophysics, meteorology, and machine learning. Early strategies relied on heuristic rules or combinatorial optimization (\cite{heaney2016validation}) to select observation locations based on expert knowledge or exhaustive search. Building upon these initial efforts, a broad spectrum of model-driven approaches has been proposed for optimal sensor placement in geophysical systems. A first category leverages the principal modes of variability derived from Empirical Orthogonal Functions (EOF) or singular value decompositions such as PCA-QR (\cite{marcille2022gaussian, manohar2018data}). These methods identify regions of maximum variance to place sensors. Working on this idea, greedy algorithms (\cite{clark2018greedy}) iteratively add sensors to maximize information gain or minimize uncertainty, typically using mutual information or variance reduction criteria. Although they can yield near-optimal configurations, their computational complexity scales poorly with the vast number of candidate grid points in the ocean. Ensemble-based sensitivity methods (\cite{hakim2020optimal}), such as ensemble transform Kalman filters, extend these principles by estimating the forecast impact of potential observations through ensemble covariance propagation. However, they still assume static observation networks and focus on short-term objectives, which limits their capacity to capture dynamically evolving oceanic structures such as eddies, filaments, and fronts. Similarly, Bayesian experimental design frameworks (\cite{alexanderian2021optimal}) state sensor placement as the maximization of expected information gain, but rely on high-dimensional Monte Carlo or variational approximations that are computationally prohibitive for non-linear, high-dimensional ocean dynamics. Even recent neural surrogate models (\cite{deng2021deep}) that approximate Bayesian objectives require extensive retraining to adapt to changing oceanic regimes. Collectively, these classical and model-based approaches suffer from limited scalability, static assumptions, and weak adaptability, motivating the transition toward data-driven, deep learning frameworks capable of learning flexible, end-to-end adaptive sensing strategies directly from data.

Recently, deep learning approaches have emerged as powerful alternatives for optimal sensor placement. The Concrete AutoEncoder (CAE) (\cite{balin2019concrete}) introduced a differentiable relaxation of discrete sensor selection through the Gumbel–Softmax trick (\cite{maddison2016concrete, jang2016categorical}), enabling gradient-based optimization of sensor locations. Initially tested on low-resolution datasets, this method has been extended to high-resolution oceanic applications (\cite{lobashevgeneralization}) and hyperspectral imaging (\cite{sun2021novel}), but it still relies on perfect ground-truth supervision, which is rarely available in operational forecasting contexts. Convolutional Gaussian Neural Processes (ConvGNPs) (\cite{andersson2023environmental}) further advanced the field by learning non-stationary, scale-dependent covariance structures directly from data. While they improve reconstruction accuracy for environmental variables such as sea surface temperature anomalies, ConvGNPs remain limited to static sensor networks, lack explicit sensor-budget control, and separate placement from reconstruction optimization. To overcome these limitations, joint sensor-placement and reconstruction frameworks have been recently explored. The PhySense model (\cite{ma2025physense}) formulates the problem as a differentiable variational inference task, assuming that the target field is conditionally Gaussian given sensor observations and locations. The model jointly optimizes a reconstruction network and the sensor positions by minimizing a differentiable proxy of the posterior uncertainty. While this represents a significant advance over static or surrogate-based designs, PhySense still produces fixed sensor configurations optimized offline and does not yet implement online or sequential adaptation. Finally, reinforcement learning and active control approaches (\cite{wang2023learning}) cast adaptive sampling as a Markov decision process, where an agent learns to move or activate sensors over time to improve forecast skill. While these methods enable dynamic reconfiguration, they typically require large training ensembles and suffer from limited generalization across different oceanic regimes.

Overall, there is a clear evolution from static strategies toward fully adaptive, data-driven sensing ones. Despite these advances, major challenges remain in achieving scalable, uncertainty-aware, and operationally feasible adaptive sampling for complex real-world dynamical systems, such as upper ocean dynamics. In this study, we aim to address these shortcomings by introducing a lightweight and interpretable framework for adaptive, reconfigurable sensor placement in the monitoring of ocean dynamics. Assuming the availability of forecast ensembles, our formulation unifies adaptive sampling and field reconstruction within a single, end-to-end differentiable optimization problem.

\section{Proposed approach}
We consider a state-space representation of the time evolution of an ocean state $\mathbf{u}(t)$ given sparse observations  $\mathbf{y}(t)$:
\begin{equation}
\left \{\begin{array}{ccl}
\label{eq: dynamical model}
    \displaystyle \frac{\partial \mathbf{u}(t)}{\partial t} &=& {\cal{M}}\left (\mathbf{u}(t) \right )+ \eta_1(t)\\
    \displaystyle \mathbf{y}(t) &=& {\cal{H}}\left (\mathbf{u}(t)\right )+ \eta_2(t)
\end{array}\right.
\end{equation}
where $\cal{M}$ represents the dynamical operator governing the system's evolution, which may involve non-linear oceanographic processes. The term $\eta_1(t)$ stands for the model error. It accounts for uncertainties in the physics-based model, typically modeled as Gaussian noise. Observations $\mathbf{y}(t)$ relate to the considered ocean state up to some observational noise $\eta_2$  that is also generally assumed to be Gaussian. The observation
Operator $\mathcal{H}$ applies to the full state $\mathbf{u}(t)$ and defines the space-time sampling pattern of the observations. 

Our goal is to optimize the observation operator $\mathcal{H}$, i.e. the spatial placement of the sensors, to maximize the reconstruction accuracy of $\mathbf{u}(t)$. In this study, we assess  the quality of the reconstruction performance using root mean squared error (RMSE). Although the rest of the study focuses on this metric, the framework can be generalized to any differentiable reconstruction objective, including mean absolute error, gradient-based or Laplacian-based losses that emphasize fronts and eddies, spectral losses that penalize unresolved spatial scales, or likelihood-based criteria when the reconstruction model is probabilistic. Non-differentiable metrics may also be reported for evaluation, but would require differentiable proxy losses during optimization. Formally, we state operator $\mathcal{H}$ as the realization of a Bernoulli random field, {\em i.e.} $\mathcal{H} \sim \mathrm{Bernoulli}\bigl(\sigma(\boldsymbol{\ell})\bigr)$ where the sampling probability at a given grid point $p$ follows a Bernoulli distribution with parameter $\sigma(\ell(p))$, $\sigma(\cdot)$ being the sigmoid activation, and $\boldsymbol{\ell}$ the logit field to be optimized. This Bernoulli formulation is naturally applied here because for each candidate grid point, we have a binary decision: it is either selected and observed or not. Under this parameterization, we define the following joint optimization problem with respect to the reconstruction parameters $\theta$ and the logit field $\boldsymbol{\ell}$:  

\begin{equation}
\min_{\boldsymbol{\ell},\theta}
\Biggl\{
  \mathbb{E}
    \Bigl\|\mathbf{u}^{(i)} \;-\; \Psi_{\theta}\bigl(\mathcal{H}(\mathbf{u}^{(i)};\boldsymbol{\ell})\bigr)\Bigr\|^2
  \;+\;
  \alpha \,\max\!\Bigl(0,\,
    \mathbb{E}\bigl\|\mathcal{H}\bigr\|_{1}
    \;-\; N_{\mathrm{budget}}
  \Bigr)
\Biggr\}.
\label{eq:minim}
\end{equation}
This optimization criterion balances the reconstruction performance and the sensor budget constraint. In practice, the sensor budget is fixed by operational constraints and denoted by $N_{\mathrm{budget}}$. We enforce this budget through a soft sparsity penalty weighted by $\alpha$. At the beginning of training, $\alpha$ is set to zero so that the mask can explore the full domain. It is then gradually increased, following the strategy of (\cite{lobashevgeneralization}), until the expected number of selected sensors remains below the prescribed budget.

Intuitively, each grid point is assigned a learnable score, or logit, which is transformed into a sampling probability through a sigmoid activation. During training, several candidate observation masks are drawn from this probability field. For each sampled mask, the corresponding sparse observations are passed to the OI reconstructor, and the reconstruction error is evaluated against the target field. Locations that repeatedly contribute to reducing the reconstruction error receive larger sampling probabilities through gradient descent, whereas locations that provide little information are progressively down-weighted. Importantly, the whole logit field is optimized simultaneously, rather than through a greedy one-by-one sensor selection.

To solve this optimization, we consider a gradient-based approach and leverage the Gumbel-Softmax trick (\cite{jang2016categorical,maddison2016concrete}) as a differentiable relaxation of the Bernoulli sampling process. A direct binary draw from the sampling probabilities would make the observation mask non-differentiable and would therefore prevent standard backpropagation through the sampling step. The use of Gumbel noise is motivated by the Gumbel-Max trick: adding independent Gumbel perturbations to the log-probabilities and taking an $\arg\max$ produces an exact categorical sample. This property is linked to the max-stability of the Gumbel distribution. The Gumbel-Softmax relaxation replaces this hard $\arg\max$ with a temperature-controlled softmax, yielding a continuous approximation of the binary mask during training. At high temperature, the mask remains smooth and exploratory; as the temperature is annealed, the samples become increasingly close to discrete sensor-selection masks.

At inference time, the final observation mask $\mathcal{H}(\mathbf{u}^{(i)})$ is obtained by thresholding the logits or by sampling according to the learned Bernoulli probabilities, and the Hadamard product between this mask and the input field yields the sampled observations. Our framework allows for the joint training of the sampling mask $\mathcal{H}$ and the parameters of the reconstruction operator $\Psi_{\theta}$. In this study, $\Psi_\theta$ is an Optimal Interpolation operator and $\theta$ denotes its spatial correlation length. Jointly optimizing the mask and $\theta$ allows the interpolation scale to adapt to the learned sampling pattern. The reported gains should therefore be interpreted as the performance of the coupled Gumbel-Softmax/OI system, rather than as a sampling pattern that is independent of the chosen reconstructor.

In our numerical experiments, we consider a  Monte-Carlo approximation of the expectation over the Bernoulli mask  through 45 independent realizations of $\mathcal{H}$. These masks are sampled at each optimization step to approximate the expectation over stochastic masks. The number of mask realizations is chosen as a compromise between gradient stability and computational cost. We use a warm-up phase where  weight $\alpha$ is set to $0$ before a smooth increase until the sensor budget constraint is full-filled. We also account for a temperature annealing schedule in the computation of the Gumbel-Softmax relaxation of the Bernoulli distribution. The training begins with a relatively high softmax temperature $T$, yielding smooth gradients, and $T$ is gradually lowered so that the sampled masks converge towards crisp binary patterns. All these numerical ingredients ensure both reliable gradient flows in the early stages and faithful enforcement of the discrete, budget-limited design at convergence. The training phase typically involves 1000 iterations with a learning rate of 0.1 using Adam optimizer for the gradient descent optimization.   

We also measured the computational cost of the proposed method. The experiments were run on a single NVIDIA L40S GPU. For one daily optimization with 1000 training iterations, the method required about $45$ s. Peak allocated GPU memory ranged from $3.24$ to $3.82$ GiB, and peak reserved memory ranged from $5.57$ to $7.37$ GiB. These results indicate that the proposed optimization remains lightweight for the considered SSH domain. The cropped Gulf Stream grid contains approximately $280 \times 280$ candidate locations, and a $0.1\%$ budget corresponds to about 78 sensors. The main cost comes from repeatedly applying OI during the Gumbel-Softmax optimization. For $K$ sensors and $P$ grid points, the OI prediction cost scales as $\mathcal{O}(KP)$, while the covariance inversion over observed points scales as $\mathcal{O}(K^3)$. Since $K \ll P$ in our sparse-observation setting, the computation remains tractable.

\begin{table}[ht]
\centering
\small
\begin{tabular}{lccc}
\hline
\textbf{Method} & \textbf{Time (min)} & \textbf{Peak allocated memory} & \textbf{Peak reserved memory} \\
\hline
GS known target 
& $0.747 \pm 0.039$ 
& $3.82$ GiB 
& $7.37$ GiB \\

GS perturbed, $\sigma=2$  
& $0.746 \pm 0.041$ 
& $3.80$ GiB 
& $7.33$ GiB \\

GS perturbed, $\sigma=5$  
& $0.739 \pm 0.038$ 
& $3.79$ GiB 
& $7.29$ GiB \\

GS perturbed, $\sigma=10$  
& $0.739 \pm 0.040$ 
& $3.79$ GiB 
& $7.36$ GiB \\

GS perturbed, $\sigma=20$  
& $0.725 \pm 0.038$ 
& $3.61$ GiB 
& $6.75$ GiB \\

GS perturbed, $\sigma=30$  
& $0.700 \pm 0.026$ 
& $3.24$ GiB 
& $5.57$ GiB \\
\hline
\end{tabular}
\caption{Computational cost of one daily Gumbel-Softmax optimization on a single NVIDIA L40S GPU. All experiments use the same Gulf Stream domain, a $0.1\%$ observation budget, 1000 training iterations, and the same optimization schedule. Memory is reported in GiB from PyTorch measurements.}
\label{tab:computing_requirements}
\end{table}
\section{Spatially-Coherent Perturbations for Ensemble Forecast Generation}\label{sec:spaccoh}

In operational cases, the complete and error-free ground truth is never available and is fundamentally unobservable. However, operational forecasting systems deliver ensembles of short-term forecasts to represent the probability density function of the states. To mimic this operational case-study, we consider a synthetically perturbed ensemble forecast $\mathcal{E}_\delta$, where $\delta$ parametrizes the displacement scale. These ensembles serve as proxies for ground truth. They are crafted using spatially-coherent perturbations that systematically displace dynamical structures (e.g., eddies, fronts) with controlled correlation lengths and temporal persistence, thereby emulating the positional uncertainties endemic to real-world forecasting systems. Training on these perturbed fields enables the model to optimize sensor placements when feature locations are inexact, mirroring the challenges of operational data assimilation.  These perturbations are intended to mimic a common forecast-error mode in mesoscale ocean prediction: the model may represent the correct type of structure, such as an eddy or front, but place it at a slightly incorrect location.

The ensemble forecasts are generated using spatially coherent random perturbations, following the idea that structured input perturbations can provide a controlled proxy for forecast uncertainty in ocean fields (\cite{roy2026spatiotemporal}). In our case, these perturbations are designed to mimic errors in the location of coherent dynamical structures, such as eddies, fronts, and meanders, arising from model biases or forecast errors. To ensure spatial coherence, nearby grid points are displaced in a similar way. A spatial correlation length $L$ is therefore introduced to control the smoothness of the displacement field and to align the perturbations with the dominant spatial scales of the ocean dynamics.

More precisely, two correlated Gaussian random fields, $\Delta x(\mathbf{r})$ and $\Delta y(\mathbf{r})$, are generated to represent zonal and meridional displacements, respectively. These fields are smoothed by convolution with a kernel of scale $L$, which enforces spatial coherence in the resulting deformation. The original grid $\mathbf{r}=(x,y)$ is then mapped to a displaced grid $\mathbf{r}' = \left(x+\Delta x(\mathbf{r}),\, y+\Delta y(\mathbf{r})\right),$ so that the main oceanic structures are shifted while preserving their spatial organization. The perturbed field $\mathbf{u}'$ is finally obtained by interpolating the original field $\mathbf{u}$ onto the displaced grid $\mathbf{r}'$ using bilinear interpolation.


\begin{figure}[!ht]
    \centering
    \includegraphics[width=0.75\textwidth]{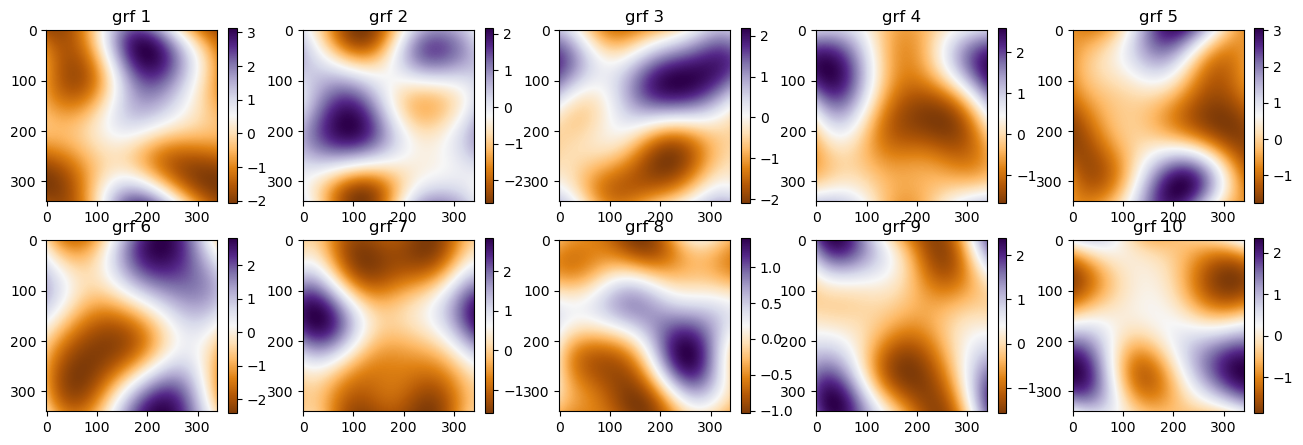}
    \caption{Example of a two-dimensional Gaussian random field used to generate spatially coherent perturbations for ensemble-forecast simulations. Each realization defines smooth displacement fields that shift oceanic structures (e.g., eddies, fronts) while preserving realistic spatial correlations.}
    \label{fig:grf}
\end{figure}

Following the same training procedure described in the previous section, the Gumbel-Softmax sampling mask $\mathcal{H}$ and the parameters of the reconstruction operator $\Psi_{\theta}$ are optimized to minimize reconstruction error on the ensemble, rather than on the unavailable ground truth. This forces the model to prioritize robustness to positional inaccuracies, for instance by assigning higher probabilities to locations along the expected migratory paths of displaced features or near regions where forecast disagreements are greatest.

\begin{equation}
        \min_{\boldsymbol{\ell},\theta}\Bigl\{\mathbb{E}_{\mathbf{u} \sim \mathcal{E}_{\delta}}\bigl[\|\,\mathbf{u} - \Psi_{\theta}\bigl(\mathcal{H}(\mathbf{u};\boldsymbol{\ell})\bigr)\|^2\bigr] \;+\; \alpha \,\max\!\Bigl(0,\,\mathbb{E}\bigl[\|\mathcal{H}\|_{0}\bigr] \;-\; N_{\mathrm{budget}}\Bigr)\Bigr\}
\label{eq:min_ens}
\end{equation}

where $\mathcal{E}_{\delta} = \{\mathbf{u}\}^{M}_{i=1}$ is the ensemble of perturbed forecasts generated by displacing ground truth structures with scale $\delta$.

\begin{figure}[!ht]
    \centering
    \includegraphics[width=0.75\textwidth]{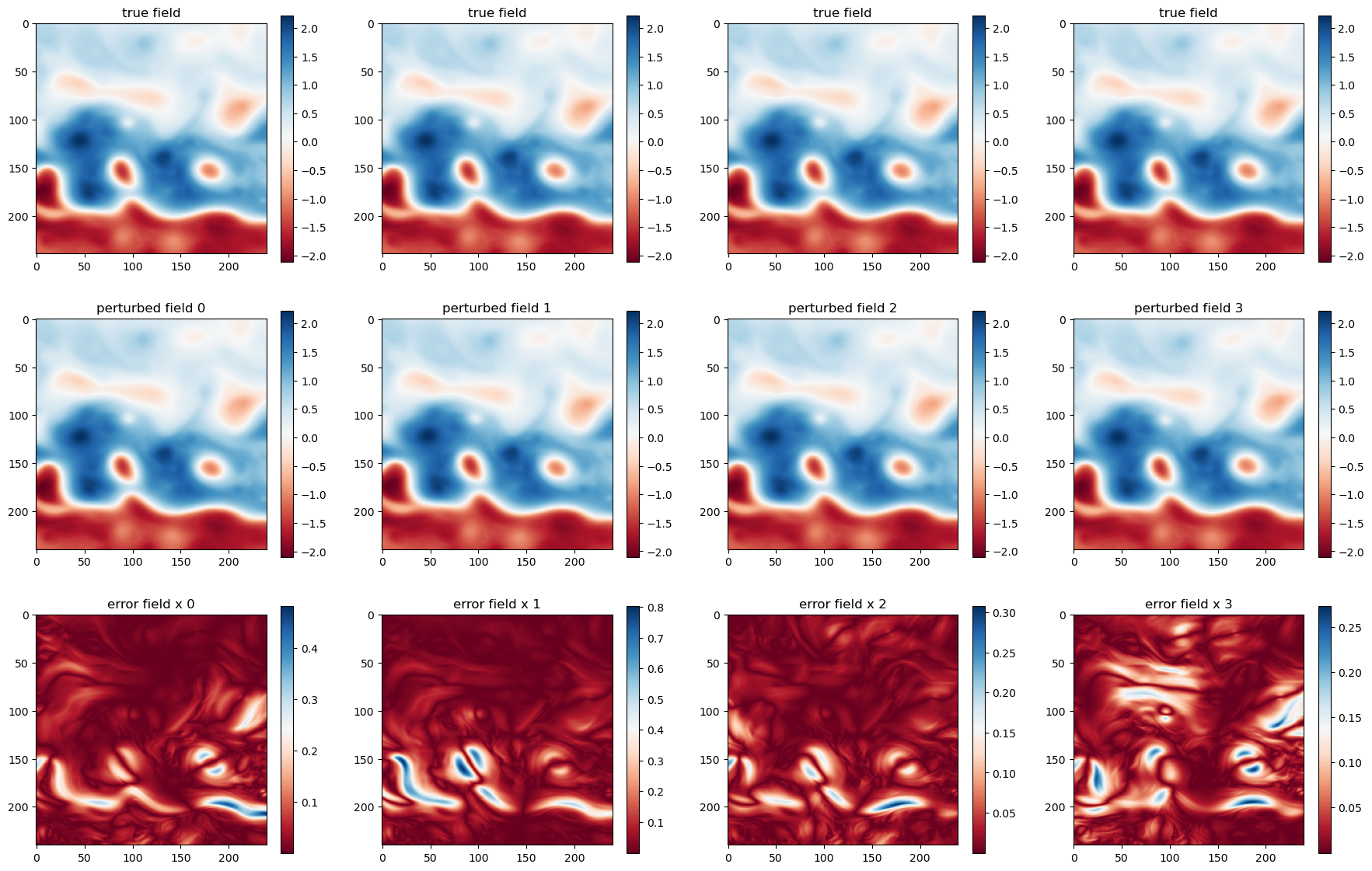}
    \caption{Illustration of the ensemble-generation process for a displacement amplitude of 1°. The original sea surface height (SSH) field (top row) is deformed using correlated Gaussian displacement fields to produce a perturbed realization (middle row) with the error field (bottom row)}
    \label{fig:grf_gene}
\end{figure}

Performance is again assessed on the ground-truth field, which the model has seen only indirectly through the synthetic ensemble during training. This evaluation measures the model's ability to recover the "true" ocean state when observational noise and model biases are absent, providing a theoretical upper bound on reconstruction fidelity. By decoupling training (on synthetic ensembles) from testing (on ground truth), we maintain that the framework remains grounded in operational realism while certifying its effectiveness under idealized conditions, a critical step for deploying adaptive systems on autonomous platforms like gliders or drifters, which must bridge the gap between imperfect forecasts and unobservable reality. Overall, this approach not only removes the need for ground truth during training but also encodes the spatiotemporal statistics of real-world errors, ensuring that sensor placements adapt to the evolving uncertainties inherent in ocean forecasting.
\section{Results}
\subsection{Dataset}
To assess the effectiveness of our approach, we conduct an Observing System Simulation Experiment (OSSE). Specifically, we exploit SSH from the NATL60 and eNATL60 numerical simulation datasets (\cite{ocean_next_2020}) obtained through the NEMO ocean model for a Gulf Stream region. Both datasets provide daily SSH fields at a $1/20^\circ$ spatial resolution. In this study, we extract the Gulf Stream region spanning $28^\circ$-$45^\circ$N and $66^\circ$-$49^\circ$W, and we remove a 30-grid-point boundary region for the optimization and evaluation. This yields an effective domain of approximately $280 \times 280$ grid points, i.e. close to a $14^\circ \times 14^\circ$ region. The NATL60 runs from October 2012 to September 2013 while the eNATL60 runs from July 2008 to June 2009.

\begin{figure}[!ht]
    \centering
    \includegraphics[width=0.4\textwidth]{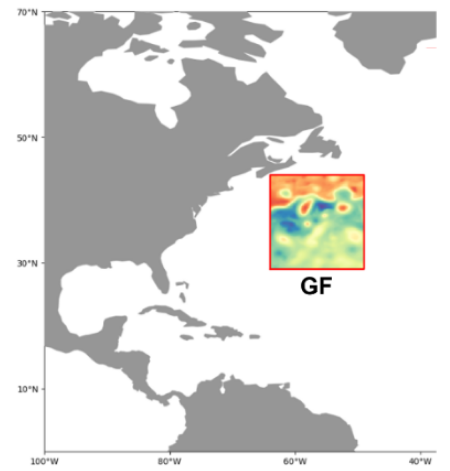}
    \caption{Geographic localization of the Gulf Stream study domain NATL60 and eNATL60 within the Gulf Stream region}
    \label{fig:GF_dataset}
\end{figure}

We compare the proposed method with two groups of baselines. The first group contains simple spatial-coverage baselines: uniform random sampling, block-random sampling, and regular-grid sampling with random jitter. The second group contains static score-based baselines commonly used in sparse sensing, including variance-based, entropy-based, EOF-energy, and PCA-QR placement \cite{manohar2018data}. Spatially constrained variants of these static baselines are also tested to reduce sensor clustering.

In this SSH/OI setting, the static score-based baselines are not competitive: they perform worse than uniform random sampling. This suggests that a fixed variance, entropy, EOF, or PCA-QR score map is not sufficient for this highly dynamic Gulf Stream region, where informative structures such as eddies, fronts, and meanders move in time. By contrast, block-random and regular-grid baselines perform better than uniform random sampling, confirming that spatial coverage is important. Nevertheless, all non-learned baselines remain clearly below the proposed Gumbel-Softmax optimization.

All the tests and methods are constrained to use only 0.1\% of the total grid points, reflecting real-world sparse operational budgets. This budget can be modified depending on the observed variable, spatial resolution, and operational constraints. For statistical robustness, for each day, 1000 observation masks are sampled independently for stochastic methods, and reconstructions are performed using Optimal Interpolation (OI) with an optimal correlation length for the baselines, while our method jointly optimizes sensor placement and OI parameters.

For each method, the baselines and the perturbed forecasts, the reconstruction quality is assessed using Root Mean Square Error (RMSE) to measure the global fidelity to the ground truth and the Explained Variance to quantify the retained spatial variability. To compare the full spectrum of outcomes, capturing not just the mean of the metrics but the standard deviation and the consistency of the methods, the metrics cumulative distribution function (CDF) is used on a 1000 independently sampled mask per method for each day. This allows us to judge whether a reconstruction is not only accurate on average but also faithful and robust across different samples.

To simulate forecast errors, we generate synthetic ensembles by displacing dynamical structures in the NATL60 fields using spatially correlated perturbations. Displacements are applied at six scales: 0° (no shift), 1/10° (~11 km), 1/4° (~27 km), 1/2° (~55 km), 1° (~110 km), and 1.5° (~165 km), covering sub-mesoscale to basin-scale uncertainties. 

\subsection{Dealing with a known target field}

\begin{table}[ht]
    \centering
    \begin{tabular}{lcc}
        \hline
        \textbf{Method} & \textbf{Mean Loss} & \textbf{Explained Variance} \\
        \hline
        Optimized mask   & 0.09078 & 0.9312 \\
        Random mask & 0.17497 & 0.7444 \\
        \hline
\end{tabular}
    \caption{Comparison of sensor placement approaches using the proposed observation mask optimization and using a purely-random (uniform distribution) strategy for sensor placement on a 100 samples for each day of the year.}
\end{table}

To assess the well-posedness of the proposed optimization problem, we first consider an idealized setting where the true target field is perfectly known, so that its distribution reduces to a Dirac delta in (\ref{eq:min_ens}). From an application standpoint, this is analogous to a compression problem where the goal is to identify an optimal subset of grid points that best preserves the spatial variability of the field. For the considered one-year dataset, we compare the reconstruction performance using  optimized sampling masks according to \ref{eq:min_ens} with $0.1\%$ of the points to the reconstruction issued from a random uniform sampling strategy. For the latter, 100 independent random masks are generated for each day to ensure statistical robustness.

The reconstruction using our method achieves a mean reconstruction loss (RMSE) of 0.09078, nearly halving the error of the random baseline (0.17497). This substantial reduction underscores the effectiveness of 
the considered gradient-based optimization in positioning observations to resolve dynamically critical regions, such as frontal zones and eddy peripheries, which are often under-sampled by uniform strategies. 

This reduction of the reconstruction error amounts to an explained variance of 93\% of the spatial variability in the ground truth field. It represents a 20\% improvement over the random baseline (0.7444). These metrics highlight the ability of the proposed method to prioritize observations in regions where positional uncertainties most degrade reconstruction performance.

\begin{figure}[!ht]
\centering
    \includegraphics[width=0.65\textwidth]{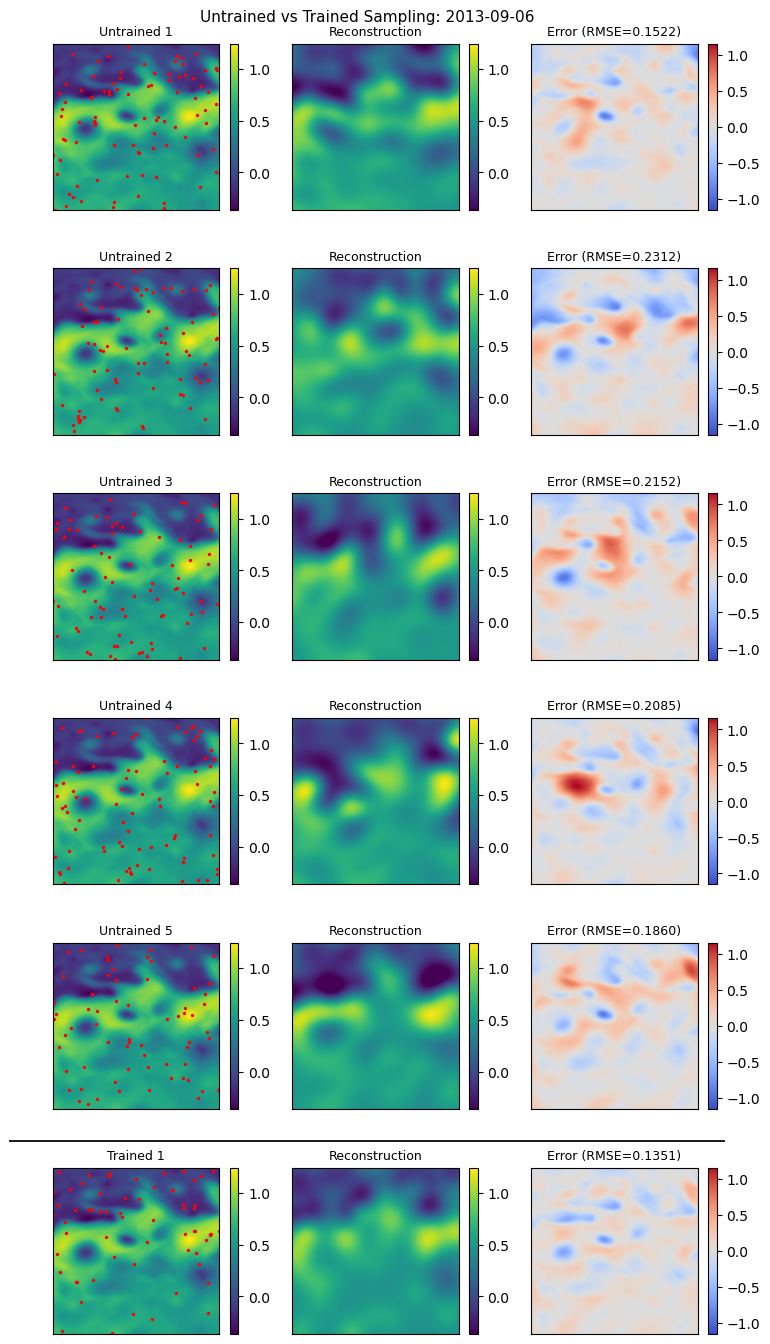}
\caption{Comparison for 2013-09-06 of the reconstruction of five untrained random strategy sampling and one trained using the Gumbel-Softmax based method. Each row shows a different sampling realization: the left panel displays the ground-truth SSH field with red dots marking selected observation locations, the middle panel presents the corresponding optimal-interpolation reconstruction, and the right panel shows the reconstruction error map with its RMSE value.}
\label{fig:sampling_reconstruction}
\end{figure}

Figure \ref{fig:sampling_reconstruction} shows the impact of the learned sampling strategy by comparing the untrained random sampling to the trained Gumbel-Softmax based one on September 6, 2013 of the eNATL dataset. The corresponding RMSE statistics in Table~\ref{tab:rmse_20130906} confirm this improvement, with the trained strategy achieve lower error and variance across realizations compared to the untrained configuration, demonstrating the effectiveness of the learned sampling distribution.

\begin{table}[ht]
    \centering
    \begin{tabular}{lcccc}
        \hline
        \textbf{Method} & \textbf{Mean RMSE} & \textbf{Std} & \textbf{Min} & \textbf{Max} \\
        \hline
        PCA-QR sampling 
        Untrained sampling & 0.1954 & 0.0240 & 0.1260 & 0.3309 \\
        Trained sampling & 0.1351 & 0.0001 & 0.1342 & 0.1357 \\
        \hline
    \end{tabular}
    \caption{RMSE statistics for 10000 realizations of the untrained and trained sampling mask on 2013-09-06.}
    \label{tab:rmse_20130906}
\end{table}

\subsection{Dealing with forecast ensembles}

We perform a second experiment where the expectation in (\ref{eq:minim}) w.r.t. target field $\mathbf{u}$ is associated with a perturbed ensemble of forecasts. As described in Section \ref{sec:spaccoh}, we aim to consider a case-study more representative of an operational setting where we are given an ensemble of forecasts to design some optimal observation strategies for the coming time window. We recall that the perturbed forecast ensembles involve random location uncertainties with a predefined horizontal range from 1/10$^\circ$ to 1.5$^\circ$. Considering an ensemble-based approximation for criterion (\ref{eq:min_ens}), we address the optimization of the sampling prior jointly with the OI parameter for each day of the considered one-year SSH time series.  

\begin{figure}[htbp]
  \centering
  \begin{minipage}{0.48\textwidth}
    \centering
    \includegraphics[width=\linewidth]{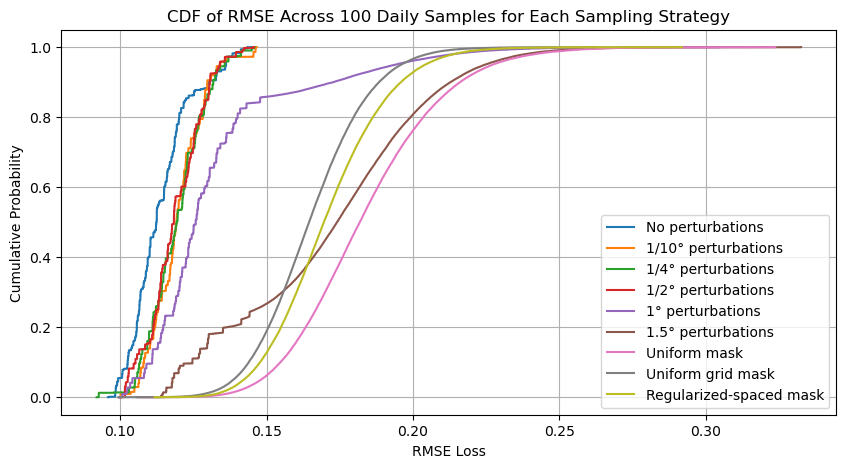}
  \end{minipage}
  \hfill
  \begin{minipage}{0.48\textwidth}
    \centering
    \includegraphics[width=\linewidth]{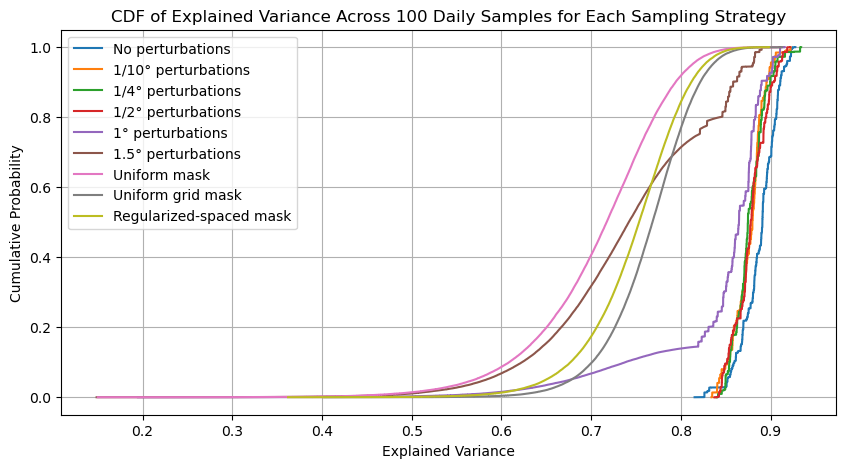}
  \end{minipage}
  \caption{Empirical CDFs of reconstruction metrics over 100 independent daily mask samples. Curves compare an adaptive Gumbel-Softmax sampling strategy, perturbed versions of the same mask displaced by increasing spatial errors (1/10°–1.5°), and random baseline mask. Metrics shown: (a) RMSE and (b) Explained Variance. All methods use the same number of sensors. Adaptive sampling yields systematically better and more robust reconstructions, though performance decreases with increasing location uncertainty.}
  \label{fig:CDF}
\end{figure}

We first analyze in Figure \ref{fig:CDF} the reconstruction performance through the empirical cumulative-distribution functions (CDF) of the RMSE and of the explained variance. We compute the CDFs over 100 samples for each day of the considered one-year time series. We compare the CDF for the proposed optimization scheme when dealing with different uncertainty levels in the forecast ensemble. As baselines, we consider three different sampling strategies. The first is a fully random sampling identical to the one used previously (uniform mask). To provide more robust comparisons, we introduce two structured alternatives: a block-stratified sampling, where the domain is divided into equal rectangular blocks and one point is randomly selected within each block, ensuring uniform spatial coverage (uniform grid mask) and a regular grid sampling where sensors are placed on a regular grid that is randomly perturbed to better capture field variability while maintaining global coverage (regularized space based mask).

\begin{figure}[!ht]
\centering
    \includegraphics[width=0.75\textwidth]{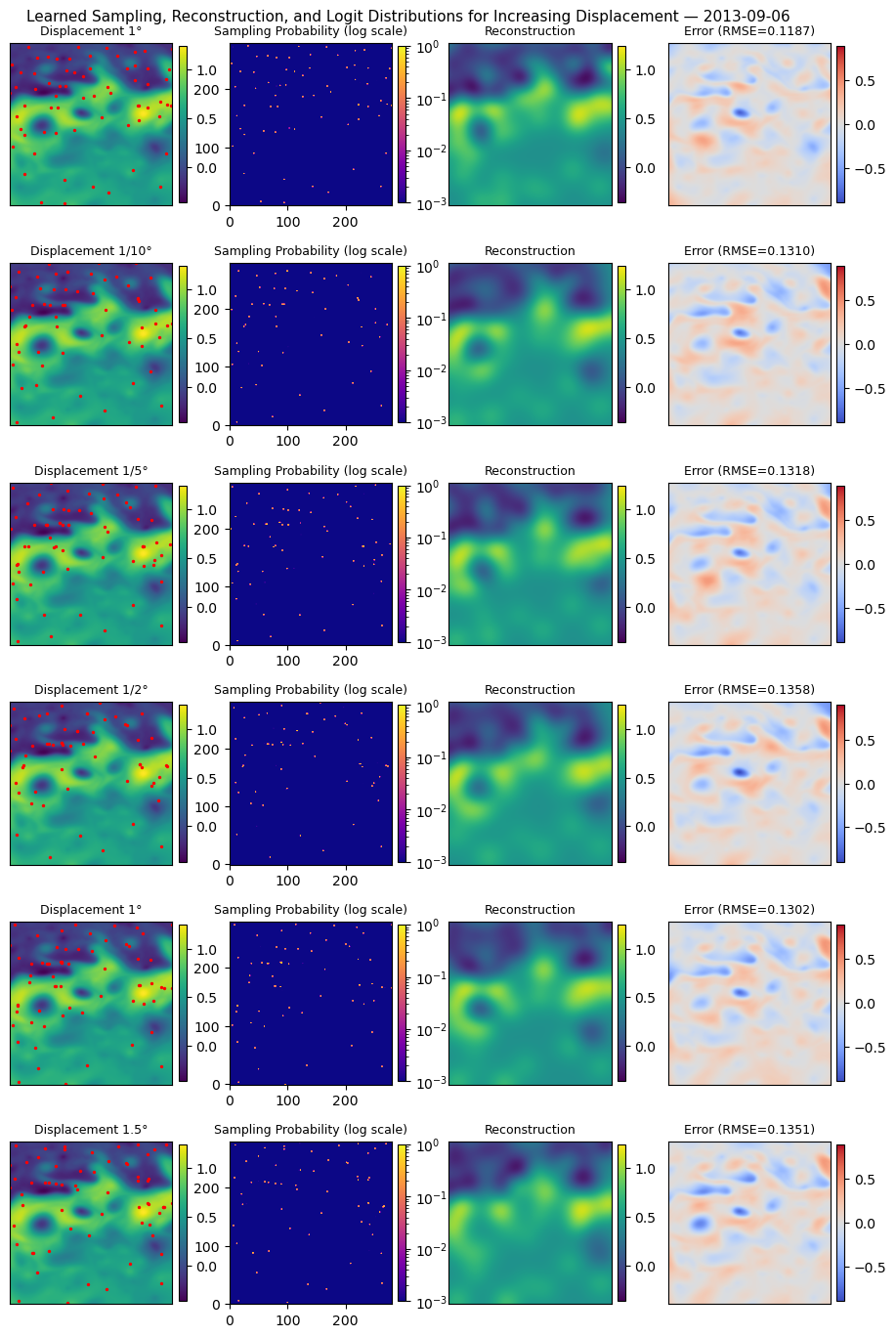}
\caption{Evolution of the learned sampling strategies and reconstruction performance for increasing displacement amplitudes. Each row corresponds to a different displacement level (from 1/10° to 1.5°), showing from left to right: the ground-truth SSH field with the optimized observation points in red, the sampling probability after training, the optimal-interpolation reconstruction, and the reconstruction error map with its RMSE value. Results are displayed for the eNATL60 dataset on 6 September 2013.}
\label{fig:displacement_sigma}
\end{figure}

Figure \ref{fig:displacement_sigma} and Table \ref{tab:rmse_sigma_20130906} illustrate the impact of increasing displacement amplitudes on the learned sampling patterns and reconstruction performance for September 6, 2013 of the eNATL60 dataset. The figure shows how the optimized observation masks adapt as the assumed displacement increases, while the table reports the corresponding quantitative reconstruction accuracy measured in terms of RMSE.
\begin{table}[ht]
    \centering
    \begin{tabular}{lcccc}
        \hline
        \textbf{Displacement level} & \textbf{Untrained Mean} & \textbf{Untrained Std} & \textbf{Trained Mean} & \textbf{Trained Std} \\
        \hline        
        0 & 0.1953 & 0.0236 & 0.1187 & 0.0001 \\
        1/10° & 0.1954 & 0.0238 & 0.1310 & 0.0001 \\
        1/4° & 0.1954 & 0.0239 & 0.1318 & 0.0004 \\
        1/2° & 0.1950 & 0.0236 & 0.1358 & 0.0002 \\
        1° & 0.1952 & 0.0236 & 0.1314 & 0.0022 \\
        1.5° & 0.1950 & 0.0235 & 0.1351 & 0.0001 \\
        \hline
    \end{tabular}
    \caption{RMSE statistics from 10,000 realizations of untrained and trained sampling strategies across different displacement levels on 2013-09-06. The trained strategies consistently outperform the untrained baseline even for high level (up to 1.5°) of displacement}
    \label{tab:rmse_sigma_20130906}
\end{table}

For up to 1/2° of forecast uncertainty, the optimized sampling captures at least 80\% of the expected variance and exceeds 85\% variance in roughly 90\% of the draws. The corresponding normalized RMSE distribution is tightly concentrated between 0.10 and 0.14. Therefore, the joint optimization does not merely improve the mean score, it also prevents catastrophic outliers. This confirms that our optimization framework which combines a warm-up phase and a temperature annealing strategy brings the optimizer to a robust optimization. It suggests that once the sensor budget is met, any further displacement field of the ensemble hardly affects skill. Taken together, these experiments support the relevance of the proposed scheme under controlled OSSE assumptions, while a full operational assessment would require real forecast ensembles and independent observations. For a 1° displacement scenario, the optimized sampling remains largely stable: nearly 95\% of the samples still outperform all unoptimized baselines, including uniform random, uniform grid, and regularized-spaced strategies. The variance CDF shows that almost all reconstructions retain more than 80\% explained variance, and the normalized RMSE remains below 0.16 for the vast majority of cases, indicating a smooth degradation rather than a breakdown. Only for the most challenging 1.5° perturbation does skill noticeably decrease, with a wider RMSE spread and reduced median explained variance; yet, even in this regime, the optimized mask remains superior to naïve baselines in roughly 70–80\% of realizations and avoids the severe tail failures observed for unoptimized strategies. These results demonstrate that the learning-based design does not simply overfit the nominal forecast ensemble; instead, it captures physically interpretable sampling structures that remain valuable even under strong spatial uncertainty, indicating that the method remains useful under the controlled forecast-displacement errors considered here.
\subsection{Analysis of the optimized sensor placement}

We analyze further the properties of the regions associated with the optimized sampling distributions. We compute the distribution of local features of the ground truth field at sampled observation points for both optimized masks and random ones.

Importantly, the Gumbel-Softmax model does not explicitly select points from a hand-coded oceanographic criterion such as SSH maxima, minima, frontal gradients, eddy centers, or regions of large forecast spread. The selected locations are learned by gradient descent through the reconstruction objective: a grid point becomes more likely to be sampled when its inclusion tends to reduce the OI reconstruction error. The following analysis should therefore be interpreted as an a posteriori diagnostic of the learned masks, rather than as an explicit selection rule imposed during training. This interpretation is consistent with sparse-sensing approaches where sensor locations are optimized for reconstruction skill and then interpreted a posteriori in terms of the underlying field structures (\cite{manohar2018data}).

\begin{figure}[!ht]
    \centering
    \includegraphics[width=0.8\textwidth]{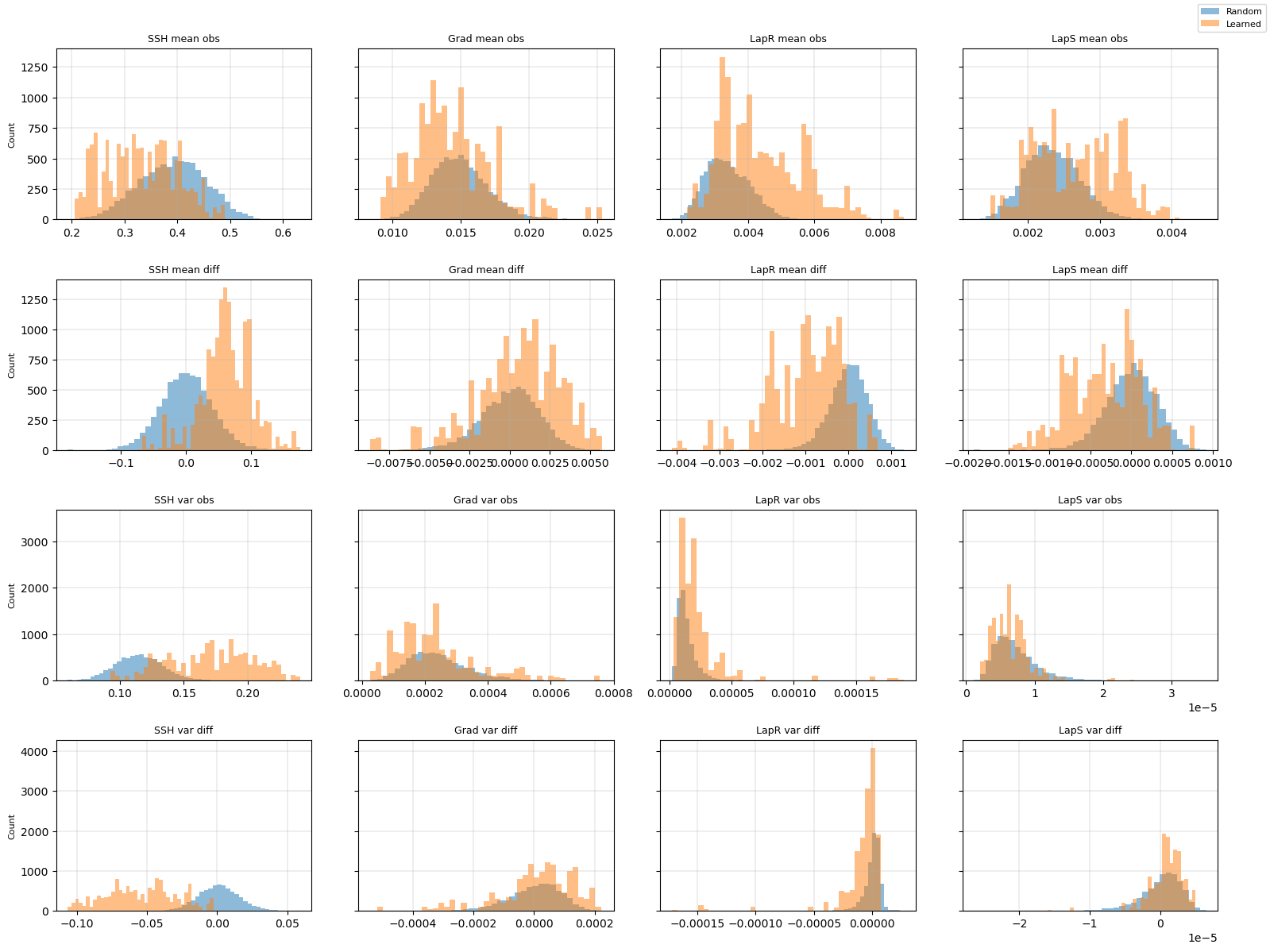}
    \caption{Scatter plots between the reconstruction metrics (RMSE and explained variance) and the sampled observation field statistic (mean and variance of the field and its gradient and Laplacian) for trained masks in red and random mask in blue.}
    \label{fig:meanvsRMSE}
\end{figure}

Figure \ref{fig:meanvsRMSE} shows that the optimized observation masks are not uniformly distributed over the SSH field, but preferentially sample locations with specific geometric properties. To quantify this behavior, we compare the values of several diagnostic fields at selected points with their distribution over the full domain. These diagnostics include SSH amplitude, gradient magnitude, signed and absolute Laplacian, local variance, and binary indicators of local maxima, minima, and extrema.

The resulting enrichment analysis indicates that the learned masks are not simply selecting large SSH amplitudes or local extrema. Local maxima, minima, and extrema are over-represented at selected locations, but their Spearman correlations with the learned sampling probability remain close to zero and their AUC values remain close to 0.5. This suggests that extrema may be selected when they are useful for reconstruction, but they are not sufficient to explain the learned sampling strategy on their own. Similarly, the gradient magnitude is not consistently enriched across all perturbation levels, so the learned mask should not be interpreted as a simple front detector.

Similarly the distribution of the Laplacian at the sampled locations shows heavier tails for the optimized masks than for purely random sampling. 
Therefore, the sampled SSH observations are more frequently associated with regions of strong local curvature. The most consistent diagnostic is the absolute Laplacian, which is enriched at selected points and shows higher top-quantile enrichment than expected from random sampling. This indicates that the optimized masks tend to favor regions where the SSH field is locally harder to interpolate, such as eddy cores, troughs, ridges, meanders, and transition zones around dynamically active structures. This behavior is consistent with the role of OI: observations in high-curvature regions provide strong constraints on the reconstructed field because such structures are more difficult to infer from distant observations than smoother regions.

The very large enrichment values obtained for the signed Laplacian should be interpreted with caution. Since the domain-average signed Laplacian may be close to zero, the corresponding selected/domain ratio can become large and sensitive to sign cancellations. For this reason, the absolute Laplacian provides a more stable and physically interpretable curvature diagnostic. Across the perturbed-forecast experiments, the curvature enrichment decreases as the displacement amplitude increases, but it remains visible for moderate perturbations. This suggests that training on displaced ensembles does not simply overfit the nominal field; instead, it preserves a preference for dynamically informative regions while becoming more robust to positional uncertainty.

In terms of variance, there is also a notable bias in the distribution. The histogram of variance values for the learned sensor locations is shifted towards higher variance. This means that the optimized masks sample points that are in regions of large fluctuation of the field while avoiding flatter and more redundant regions. However, this enrichment should again be understood as a consequence of the reconstruction objective rather than as a prescribed sampling rule. Overall, the optimized placement is best interpreted as reconstruction-driven and physically interpretable: the learned masks preferentially sample regions that are difficult for OI to reconstruct from sparse observations, especially regions of strong curvature and, to a lesser extent, local extrema and local variability. These regions often correspond to oceanographic structures such as eddies, meanders, and fronts, but the method does not explicitly detect these structures. Instead, such features emerge indirectly because they carry high information content for reconstructing the SSH field under a limited observation budget.

\section{Conclusion and Discussion}

This work bridges the gap between oceanographic data assimilation and machine learning, offering a scalable solution for observational network design under resource constraints. By focusing on sea surface height (SSH) as a test case, we highlighted the broader applicability of the framework to other spatially correlated geophysical fields while keeping the present numerical validation restricted to a controlled SSH OSSE.

We have presented a new end-to-end framework that iteratively optimizes a probabilistic sensor sampling mask to determine the best sensor placement that minimizes the reconstruction RMSE. The sampling mask is seen as a Bernoulli random field with learnable logits corresponding to location-specific sampling scores and is paired with the simple but powerful OI with a learnable correlation length. For each iteration, a Monte-Carlo estimate drawn on multiple mask realizations approximate approximates the expected reconstruction error, while a combination of a budget-free warm-up phase and a temperature annealing schedule promotes exploration at the beginning of training and progressively drives the mask toward a more discrete, budget-constrained solution.

Our preliminary results in an ideal setting where the target field is fully known show substantial gains: nearly halving the SSH reconstruction RMSE and increasing the explained variance by about 20\% compared to a traditional random sampling baseline. These improvements underscore that jointly learning the sampling pattern and reconstruction parameters can substantially improve over static or heuristic sensor layouts of the same size. In essence, by co-designing “what” to measure and “how” to reconstruct, the framework extracts far more value from each observation than uniformly random placement.

The baseline experiments clarify what the proposed method is improving over. Static score-based strategies such as variance-, entropy-, EOF-energy-, and PCA-QR-based placement are standard sparse-sensing references (\cite{manohar2018data}), but they are not effective in our SSH/OI experiments: they perform worse than uniform random sampling. This negative result suggests that, in a highly dynamic Gulf Stream region, fixed climatological or modal score maps do not provide a reliable proxy for the information needed by OI reconstruction.

This approach has several notable strengths. First, our framework optimizes directly on the field the quantity that matters to operational users, namely the mean-squared error of the reconstructed physical field, instead of a proxy such as PCA variance or information entropy. Most existing approaches optimize those surrogate criteria (\cite{marcille2022gaussian, manohar2018data, andersson2023environmental}); by contrast, our loss is fully aligned with the end goal. Because the design is tied to the reconstruction objective, it can in principle be re-optimized when new forecast ensembles become available. In the present study, however, this capability is demonstrated only in an OSSE setting. Operational deployment would require further validation with real forecast ensembles, platform constraints, and independent observations.

Secondly, we tested the method's robustness under conditions that partially mimic one important source of operational forecast uncertainty. We generated an ensemble of synthetic forecast proxies by applying controlled spatial shifts to the ground-truth field. Even when the optimization was carried out on these perturbed ensembles with displacement as much as 1$^\circ$, our method still outperformed all baselines by a comfortable margin. This resilience to positional errors indicates that the learned sensor layout does not merely overfit the nominal field in the controlled experiments. However, these perturbations represent only one class of model--reality mismatch: spatial displacement of coherent structures. Real operational errors may also involve amplitude biases, missing processes, unresolved scales, atmospheric-forcing errors, boundary-condition errors, and observation errors.

Furthermore, the masks learned through our method can be interpreted. The selected-point analysis shows that the learned mask is not equivalent to a hand-crafted detector of SSH extrema, fronts, or high-amplitude regions. The strongest and most consistent enrichment is associated with curvature diagnostics, especially the absolute Laplacian, whereas gradient magnitude and SSH amplitude are not systematically enriched across all perturbation levels. Local extrema are over-represented at selected locations, but their low predictive power when considered alone indicates that they are not sufficient to explain the learned mask. We therefore interpret the optimized placement as reconstruction-driven: the method tends to select regions that are difficult for OI to infer from surrounding observations, such as eddy cores, ridges, troughs, meanders, and transition zones. These oceanographic structures are not explicitly imposed during training; they emerge because they carry high information content for reconstructing the SSH field under a limited observation budget.

The learned probability field can also be converted into a fixed observing network by selecting the $N_{\mathrm{budget}}$ grid points with the largest sampling probabilities. Thus, although the training procedure is stochastic, the resulting deployment does not need to be stochastic. This deterministic top-budget mask is relevant for fixed observing systems such as moorings or repeated in situ sampling locations, whereas stochastic sampling is mainly used here to characterize robustness and variability across possible masks.

At the same time, some limitations remain. The performance of our framework depends on the realism and representativeness of the synthetic ensemble used for training. If the training ensemble fails to capture certain bias or error patterns present in real forecasts, the resulting sensor placement could be suboptimal when deployed in the real world. A deeper investigation using more diverse or higher-fidelity forecast ensembles is needed before operational implementation. Additionally, the need to use Monte Carlo sampling to handle the mask’s stochasticity increases the per-iteration computational cost. Therefore, possible extensions include developing more sophisticated ensemble-generation techniques , exploring different covariance or kernel parameterizations for the reconstruction model, and implementing variance-reduction strategies for the gradient estimation. Such improvements could further enhance the method’s practicality and efficiency.

Although this study focuses on SSH, the formulation extends naturally to time-dependent multi-variable geophysical states. The state vector may be written as $\mathbf{u}(t)=(\mathbf{u}_1(t),\ldots,\mathbf{u}_K(t))$, where each component corresponds to a different variable such as SSH, SST, salinity, chlorophyll, or subsurface temperature. The reconstruction loss can then be defined as a weighted sum of normalized variable-wise errors, possibly augmented with temporal regularization or cross-variable consistency terms. Depending on the observing platform, the learned sampling mask may be shared across variables, time-dependent, or optimized separately for each observing modality.

In summary, our results show that jointly learning where to observe and how to reconstruct can provide substantial and consistent gains over random and structured heuristic sampling under the controlled OSSE assumptions considered here, without increasing the sensor budget. The framework offers a flexible and interpretable step toward budget-aware observation-network design, but its operational use should be assessed with realistic forecast ensembles, independent observations, deployment constraints, and multi-variable extensions.
\begin{Backmatter}

\paragraph{Acknowledgments}
This study benefited from HPC and GPU resources from GENCI-IDRIS (Grant 2021-101030) and by the Région Bretagne, via the project CPER AIDA (2021–2027).

\paragraph{Funding Statement}
This study was carried out within the framework of the FASCINATION project funded by DGA and led by Shom. It was also supported by ANR through AI chair OceaniX. 

\paragraph{Competing Interests}
 The authors declare none.

\paragraph{Data Availability Statement}
 The code for the algorithm is available at https://github.com/CIA-Oceanix/GumbleSoftmaxOptimalPlacement. The data that support the findings of this study are available from https://github.com/ocean-next/eNATL60

\paragraph{Ethical Standards}
The research meets all ethical guidelines, including adherence to the legal requirements of the study country.

\paragraph{Author Contributions}
Conceptualization: O.C; R.F.; Y.S Methodology: O.C; R.F. Data visualisation: O.C. Writing original draft: O.C; R.F. All authors approved the final submitted draft.


\printbibliography

\end{Backmatter}
\end{document}